\documentclass[prb,onecolumn,nofootinbib,citeautoscript,10pt,notitlepage]{revtex4-2}
\usepackage{graphicx,bbold,float,slashed}
\usepackage{dcolumn}
\usepackage{color}
\usepackage{amssymb,amsmath}
\usepackage{tabularx,graphicx}
\usepackage{epstopdf}
\usepackage{latexsym}
\usepackage{colortbl}
\usepackage{psfrag}
\usepackage{bbm,bm,array,physics}
\usepackage{titlesec}
\usepackage{dsfont}
\usepackage[tight]{subfigure}

\usepackage[papersize={8.5in,11in}]{geometry}

\definecolor{darkblue}{rgb}{0.,0.,0.4}
\definecolor{darkred}{rgb}{0.5,0.,0.}
\definecolor{BlueViolet}{RGB}{138,43,226}
\definecolor{SkyBlue}{RGB}{30,144,255}
\definecolor{DarkGreen}{RGB}{0,100,0}
\usepackage[pdftex,colorlinks=true,linkcolor=darkblue,
citecolor=blue,urlcolor=darkred]{hyperref}
\def \be{\begin{align}}
\def \ee{\end{align}}
\def \bea{\begin{eqnarray}}
\def \eea{\end{eqnarray}}
\def \nn{\nonumber \\}
\renewcommand{\vec}[1]{{\mathbf{#1}}}
\begin{document}

\title{Stable non-Fermi liquid fixed point at the onset of incommensurate $2k_F$ charge density wave order}

\author{Ipsita Mandal}
\email{ipsita.mandal@snu.edu.in}

\affiliation{Department of Physics, Shiv Nadar Institution of Eminence (SNIoE), Gautam Buddha Nagar, Uttar Pradesh 201314, India
\\and\\
Freiburg Institute for Advanced Studies (FRIAS), University of Freiburg, D-79104 Freiburg, Germany}

\begin{abstract}
We consider the emergence of a non-Fermi liquid fixed point in a two-dimensional metal, at the onset of a quantum phase transition from a Fermi liquid state to an incommensurate charge density wave (CDW) ordered phase. The momentum of the CDW boson is centred at the wavevector $ \boldsymbol{ \mathcal Q} $, which connects a single pair of antipodal points on the Fermi surface with antiparallel tangent vectors. We employ the dimensional regularization technique in which the co-dimension of the Fermi surface is extended to a generic value, while keeping the dimension of the Fermi surface itself fixed at one. Although the system is strongly coupled at dimension $d=2$, the interactions become marginal at the upper critical dimension $d=d_c$, whose value is found to be $ 5/2$. Using a controlled perturbative expansion in the parameter $\epsilon = d_c - d $, we compute the critical exponents of the stable infrared fixed point characterizing the quantum critical point. The scalings of the original theory are determined by setting $\epsilon =1/2$, where the fermion self-energy is seen to scale with the frequency with a fractional power law of $2/3$, which is the telltale signature of a typical non-Fermi liquid phase.
\end{abstract}

\maketitle

\tableofcontents

%%%%%%%%%%%%%%%%%%%%%%%%%%%%%%%%%%%%%%%
\section{Introduction}

While Landau's Fermi liquid theory has been incredibly successful in describing normal metals, there exists an extensive number of metallic states where the framework fails. Although of widely different origins, these systems are widely known as non-Fermi liquids. A finite density of nonrelativistic fermions interacting with transverse $U(1)$ gauge field bosons was the first model, considered by Holstein \textit{et al.} \cite{holstein} to study the effects of the electromagnetic fields on a metal, to exhibit a non-Fermi liquid character. Subsequently, it was realized that similar behaviour is found for finite-density fermions coupled with massless order parameter bosons at quantum critical points \cite{max-isn, ogankivfr, metzner, delanna, kee, lawler1, rech, wolfle, maslov, quintanilla, yamase1, yamase2, halboth,
jakub, zacharias, eaKim, huh, denis, ips-uv-ir1, ips-uv-ir2, ips-subir, ips-sc, ips-sc_err, max-sdw, chubukov1, Chubukov, shouvik2, ips-c2, andres1, andres2} or artificial gauge field(s) emerging in various kinds of scenarios \cite{baskaran,larkin, PhysRevLett.63.680,leenag, blok, ubbens, nayak1,Chakravarty}.
While the Landau quasiparticles provide a natural single-particle basis for writing down the corresponding low-energy quantum field theories (QFTs) on the merit of being long-lived excitations in a Landau Fermi liquid, they are destroyed by the strong interactions between the soft fluctuations of the Fermi surface and the gapless bosonic quantum fields in the scenarios described above. Since these are fundamentally strongly-interacting theories, it is a challenging task to build a controlled approximation amenable to theoretical analysis. As a result, there have been intensive efforts to devise QFT frameworks to explain the emergent physical characteristics of these non-Fermi liquid systems \cite{holstein, reizer,leenag, HALPERIN,polchinski,ALTSHULER,Chakravarty,eaKim,nayak,nayak1,lawler1,SSLee,
max-sdw,max-isn, chubukov1, Chubukov,mross,Jiang, ips2, ips3,Shouvik1,denis,shouvik2, ips-uv-ir1, ips-uv-ir2, ips-subir,ips-sc, ips-sc_err, ips-c2, andres1, andres2, Lee_2018, ips-fflo, ips-nfl-u1, ips-rafael}. In two spatial dimensions, the corresponding theories are genuinely strongly interacting, while in three dimensions, they emerge as marginal Fermi liquids \cite{ips-uv-ir1, ips-nfl-u1}.
%%%%%%%%%%%%%%%%%%%%%%%%%%%%%%%
Analogous situations arise in two-dimensional (2d) and three-dimensional (3d) semimetals, where a non-Fermi liquid state emerges when the chemical potential cuts a band-crossing point giving rise to a Fermi point (rather than a Fermi surface) and a long-ranged (i.e., unscreened) Coulomb potential is switched on \cite{abrikosov,moon-xu, rahul-sid, ips-rahul, ips-qbt-sc, ips-hermann, ips-hermann2, ips-hermann3, juricic, ips-birefringent, ips-hermann-review}.

For the non-Fermi liquids arising at quantum critical points, there are two types of order parameter bosons: (1) cases where the critical bosonic field is centred about zero momentum, inducing the quasiparticles to lose coherence across the entire Fermi surface \cite{max-isn,ogankivfr,metzner,delanna,kee,lawler1,rech,wolfle,maslov, quintanilla,yamase1,yamase2,halboth,
jakub,zacharias, eaKim, huh,denis, ips-uv-ir1,ips-uv-ir2, ips-subir,ips-sc, ips-sc_err}; (2) the momentum of the quantum field describing the bosonic degrees freedom is centred around a finite wavevector equal to $  \boldsymbol{ \mathcal Q} $, which connects points on the Fermi surface (commonly known as \textit{hot-spots}), such that the non-Fermi liquid behaviour emerges locally in the vicinity of the hot-spots \cite{max-isn,chubukov1,Chubukov,shouvik2, ips-c2, andres1, andres2, ips-fflo}. Examples in the first category include the Ising-nematic critical point, while the second one include ordering transitions to phases like spin density wave (SDW), charge density wave (CDW) \cite{max-sdw, chubukov1,Chubukov,shouvik2,ips-c2,andres1,andres2},
and the Fulde-Ferrell-Larkin-Ovchinnikov (FFLO) states \cite{ips-fflo}.

The CDW (SDW) bosons with $ \boldsymbol{ \mathcal Q} \neq 0$ give rise to instabilitites involving charge (magnetic) order, where the charge (spin) density spontaneously breaks translational symmetries and develops a density modulation equal to $ \boldsymbol{ \mathcal Q} \neq 0$.
There are two distinct categorizations depending on the nature of the wavevector: (1) whether it is commensurate or incommensurate; (2) whether $ \boldsymbol{ \mathcal Q} $ is a nesting vector of the Fermi surface or not. While the so-called commensurate wavevectors can be written as a linear combination $\mathbf G$ of the reciprocal lattice vectors $\lbrace \boldsymbol {\mathcal R} \rbrace $ with rational coefficients, we cannot do the same for the incommensurate wavevectors. Clearly, there are infinitely many possible rational coefficients, but the quantitative effects of commensurability decrease with the size of their denominators. For the special situation of $ \boldsymbol{ \mathcal Q} $ equalling a nesting vector connecting two points on the Fermi surface with antiparallel Fermi velocities (or, equivalently, tangent vectors), the spin and charge orderings feature a well-known singularity caused by an enhanced phase space for low-energy particle-hole excitations. In an inversion-symmetric crystal with the valence band dispersion $\xi (\mathbf k)$, the nesting vectors $ \boldsymbol{ \mathcal Q}$ are given by the condition $\xi ( \boldsymbol{ \mathcal Q}/2  +\mathbf G/2) = \xi_{k_F} $, where $\xi _{k_F}$ is the Fermi energy. The nesting-vector nature of $ \boldsymbol{ \mathcal Q} $, coupled with inversion symmetry, implies that $| \boldsymbol{ \mathcal Q}  | =2\, k_F$, where $k_F $ is the local radius of curvature/reciprocal of curvature of the Fermi surface (i.e., the magnitude of the local Fermi momentum vector). This results from the fact that the two hot-spots are related by inversion symmetry and, hence, both have the same value of $ k_F $.
The $2k_F$-wavevector instabilities are ubiquitous in 2d systems exhibiting high-temperature superconductivity --- for example, (1) the ground state of the 2d Hubbard model at half-filling of the conduction band exhibits an SDW instability at a $2k_F$-wavevector \cite{hubbard, hubbard2}; (2) d-wave bond charge order, triggered by antiferromagnetic fluctuations, in models for cuprate superconductors occurs naturally at $2k_F$-wavevectors \cite{max-sdw, bond-order}.
We would like to emphasize that such a nesting vector causes a partial nesting of the Fermi surface, which is distinct from the perfect-nesting cases when slices (and not just discrete points) of a Fermi surface are connected by the same nesting vector. 
% PHYSICAL REVIEW B 97, 155159 (2018)  https://journals.aps.org/prb/pdf/10.1103/PhysRevB.97.155159
The evidence of incommensurate CDW orderings has been reported in materials like NbSe$_2$ and TaS$_2$ \cite{salvo, Scholz}, VSe$_2$ \cite{pai}, SmTe$_3$ \cite{Gweon}, and TbTe$_3$ \cite{Kapitulnik}. In some of these compounds, the CDW transition temperature can be tuned close to zero Kelvin by applying high pressure, unravelling a putative quantum critical point at the onset of the CDW ordering \cite{littlewood}. Such observations indicate the importance of a theoretical understanding of quantum critical points involving incommensurate $2k_F$-wavevector instabilities.

In this paper, we consider a pair of antipodal points on a 1d Fermi surface (of a 2d metal) with parallel tangent vectors interacting with an order parameter boson, whose condensation gives rise to an incommensurate CDW ordered phase \cite{metzner1, metzner2, matthias_2kf}. The CDW boson becomes massless right at the quantum critical point, giving rise to strong quantum fluctuations. The physical picture here is that the CDW boson drives the system across a quantum phase transition to an ordered state, where the electron density spontaneously breaks translational symmetries and develops a density modulation with a wave vector $  \boldsymbol{ \mathcal Q} $ that is incommensurate with the underlying reciprocal lattice vectors. For the sake of definiteness, we choose $  \boldsymbol{ \mathcal Q} = 2\, k_F \,{\hat{\boldsymbol x}}$, without any loss of generality.

To deal with the QFT describing the above system, we implement the analytic approach of dimensional regularization \cite{senshank, denis, ips-uv-ir1, ips-uv-ir2, ips-sc, ips-subir, ips-fflo, ips-nfl-u1}, in which the co-dimension of the Fermi surface is increased (as a mathematical tool) with the aim to obtain the value of the upper critical dimension $d=d_c$. Since $d_c$ is the dimension at which the interactions become marginal, we succeed in formulating a controlled perturbative approximation, although the quasiparticle-description has broken down. The critical exponents and various physical properties can now be calculated in a systematic expansion involving the perturbative parameter $\epsilon = d_c -2 $. Halbinger \textit{et al.} \cite{matthias_2kf} have considered this incommensurate CDW problem by employing the same methodology, and have found that the fermion-boson interactions lead to a stable non-Fermi liquid fixed point. Furthermore, their results show that the resulting critical Fermi surface is flattened at the hot-spots. However, they arrived at their conclusions from some inaccurate computations, which, in addition to other results, could not find the correct critical dimension $d_c = 5/2$. They carried out their computations assuming that $d_c$ is equal to $5/2$, predicting that it would be obtained at the two-loop order. Consequently, their answers do not include the all-important frequency depedence of $\text{sgn} (k_{0}) |k_{0}|^{2/3} $ for the fermion self-energy, which is predicted in the earlier (uncontrolled) random phase approximation
(RPA) calculations \cite{metzner1,metzner2}. In order to address these inadequacies, the analysis via the dimensional regularization scheme must be reexamined, if only to put it on a firmer basis.

The paper is organized as follows. In Sec.~\ref{secmodel}, we introduce the effective low-energy Euclidean action in the Matsubara frequency space, describing the fermionic excitations at the two hot-spots of the Fermi surface interacting with the CDW boson fluctuations. The original theory is embedded in $d=2$ spatial dimensions. We also explain how to generalize the theory to a generic value of $d$ by increasing the number of dimensions perpendicular to the 1d Fermi surface, as this will allow us to identify the upper critical dimension $d_c$ and, subsequently, to regularize the theory via the renormalization group (RG) procedure. Sec.~\ref{secselfen} is devoted to the derivations of the bosonic and fermionic self-energies, and showing that $d_c$ comes out to be $5/2$. Using the one-loop results, the RG flow equations are determined in Sec.~\ref{secrg} by absorbing the ultraviolet (UV) divergences into singular counterterms. We discuss the nature of the fixed points in the infrared (IR) limit and show that the system flows to a stable non-Fermi liquid point. Finally, we conclude with the relevant discussions in Sec.~\ref{secsum}, comparing our computations and results with earlier works. The appendix shows a part of the computations of the fermion self-energy.

%%%%%%%%%%%%%%%%%%%%%%%%%%%%%%%%%%%%%%%%%
\section{Model}
\label{secmodel}

%%%%%%%%%%%%%%%%%%%
\begin{figure}[t]
\begin{center}
\includegraphics[width = 0.35 \textwidth]{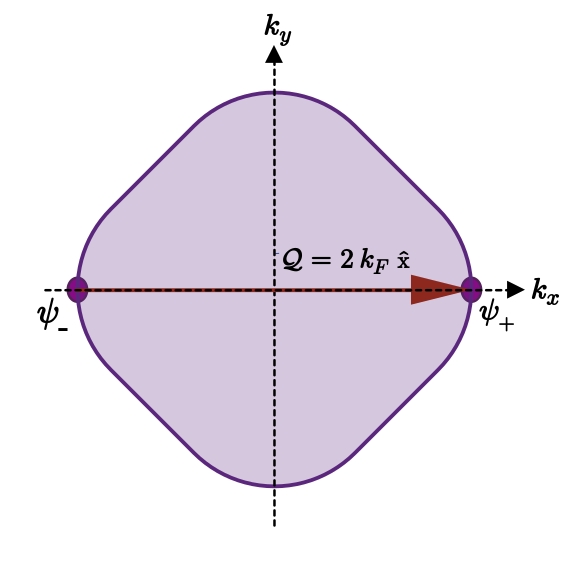} 
\end{center}
\caption{Schematics of the one-dimensional Fermi surface with two hot-spots connected by the wavevector $ \boldsymbol{\mathcal Q} = 2 \,k_F\,\hat{\boldsymbol x} $ (indicated by the red arrow), which is incommensurate with the underlying reciprocal lattice vectors. The fermionic fields in the vicinity of the right and left hot-spots are designated as $\psi_+$ and $\psi_-$, respectively. They interact with the CDW order parameter bosonic fields whose momenta are centred at $\boldsymbol{\mathcal Q}$.
\label{fig_fs}}
\end{figure}
%%%%%%%%%%%%%%%%%%%%%%%%%%%%%%%

We consider the low-energy QFT action describing finite-density fermions confined to two spatial dimensions, and interacting with an incommensurate CDW order parameter with momentum centred at $ \boldsymbol{ \mathcal Q} = 2 \,k_F\,\hat{\boldsymbol x}$. Thus, the nesting vector $  \boldsymbol{ \mathcal Q} $ connects two hot-spots on the Fermi surface located along the $x$-axis, as shown in Fig.~\ref{fig_fs}.
In $(2+1)$-dimensions, the effective action describing the electrons near the hot-spots and the CDW order parameter mode is given by \cite{matthias_2kf, metzner1, metzner2}
\begin{align}
 S &= \sum_{s=\pm}  \int_{k} \psi_{s}^{\dagger}(k)
  \left( -i\, k_0 + s \,k_1 + k_2^2 \right) \psi_{s}(k) 
	+ \int_k \phi_+(k) \left( k_0^2 + k_1^2 + k_2^2 \right) \phi_-(-k) \nn
%%%%%%%
& \qquad + e  \int_{k,q} \,
\Big[ \,\phi_+(q) \,\psi^{\dagger}_{+} (k+q) \,\psi_{-}(k) 
+ \phi_-(-q) \,\psi^{\dagger}_{-}(k-q)\, \psi_{+}(k) \,\Big]\,,
\label{eqs0}
\end{align}
where $k=(k_0,\mathbf k)$ denotes the three-vector comprising the Matsubara space frequency $k_0 $ and the spatial momentum vector $\mathbf k =(k_1, k_2) \equiv (k_x, k_y)$, $\int_k \equiv \int dk_0\, d^d{\mathbf k} /(2\,\pi)^{d+1} $, and $d =2 $ is the number of spatial dimensions . The fermionic degrees of freedom about the right and left hot-spots are denoted by $\psi_+(k)$ and $\psi_-(k)$, respectively. The fields $\phi_+(k)$ and $\phi_-(k)$ refer to the bosonic fluctuations carrying frequency $k_0$ and momenta $  \boldsymbol{ \mathcal Q}  + \mathbf k$ and $-  \boldsymbol{ \mathcal Q} + \mathbf k$, respectively. 
The bosons are massless as we are considering the quantum critical point.
To simplify notations, we have rescaled the fermionic momenta in a way such that the absolute value of the Fermi velocity is unity and the curvature of the Fermi surface is equal to $2$ at the hot-spots. Although the bosonic velocity is in general distinct from that of the fermions, we have set the bare velocity of the bosons equal to unity as well. This is because the dynamics of the bosons at the critical point is dominated by the particle-hole excitations of the Fermi surface at low energies, and the actual value of the bosonic velocity does not matter in the low-energy effective theory.

In our action, since the Fermi surface is locally parabolic, we can set the scaling dimensions of $k_1$ and $k_2$ are equal to $1$ and $1/2$, respectively.
In order to extract the critical scalings in a controlled approximation,
we increase the co-dimensions of the Fermi surface \cite{senshank, denis, shouvik2} to eventually determine the upper critical dimension $d=d_c$, where the fermion self-energy shows a logarithmic singularity.
To preserve the analyticity of the theory in momentum space 
(alternatively, locality in real space) with generic co-dimensions, 
we introduce the two-component ``spinors'' \cite{denis,ips-uv-ir1,ips-uv-ir2,ips-fflo,ips-nfl-u1,ips-rafael}
\begin{align}
 \Psi (k) = \left( 
\psi_{+}(k)\quad
\psi_{-}^\dagger(-k)
\right)^T \text{ and } \bar \Psi \equiv \Psi^\dagger \,\gamma_0\,,
\end{align} 
and write an action that describes the 1d Fermi surface
embedded in a $d$-dimensional momentum space:
\begin{align}
\label{eqs1}
S  &=    \int_k \bar \Psi(k) \,i
\left(  \vec \Gamma \cdot \vec K  +  \gamma_{d-1} \, \delta_k \right )
 \Psi(k)  
 +
 \int_k 
\left ( k_d^2 +  
\tilde a\, e_k \right ) \,  \phi_+ (k) \, \phi_-(-k) 
 \nn & \quad 
- \frac{ i\, e \, \mu^{x_e/2} } {2} 
\int_{k, \,q}
\Big[ \,\phi_+(q) \,
 \bar{\Psi} (k+q) \, \gamma_0 \, \bar{\Psi}^T(-k) 
- \phi_-(-q) \,\Psi^T(q-k) \,\gamma_0 \,\Psi(k) \Big] \,,
%%%%%
\nn x_e &=  \frac{ 5 } {2} - d \,, \quad 
\delta_k = k_{d-1} + k_d^2
\,, \quad
e_k = k_{d-1} +\frac{k_d^2}{2} \,.
\end{align}
%%%%%%%%%%%%%%%%%
The $(d-1)$-component vector $\vec K ~\equiv ~(k_0, k_1,\ldots, k_{d-2})$ includes
the frequency and the $(d-2)$-components 
of the momentum vector due to the added co-dimensions. The original momentum components along the $x$- and $y$-directions have been relabelled as $k_{d-1}$ and $k_d$, respectively.
Hence, in the resulting $d$-dimensional momentum space, the set of components
$ \lbrace k_1, \cdots ,k_{d-1} \rbrace $ represents the $(d-1)$ directions perpendicular to the Fermi surface, while $k_d$ is along the parallel direction. Similarly, the vector of matrices $\vec \Gamma \equiv (\gamma_0, \gamma_1,\ldots, \gamma_{d-2})$ has $(d-1)$ components
representing the gamma matrices associated with $k_0$ and the extra co-dimensions.
Ultimately, we are interested in continuing to $d=2$, which implies that, in practice, 
it is sufficient to consider only the $2 \times 2$ gamma matrices
$\gamma_0 = \sigma_y $ and $ \gamma_{d-1} = \sigma_x$ in our computations.

In the purely bosonic part of the action, only the $k_d^2 $ part of the kinetic term
is retained, because $\left( |\vec K|^2 + k_{d-1}^2 \right) $ 
is irrelevant under the scaling of the patch-theory formalism \cite{max-isn,denis,ips-uv-ir1,ips-uv-ir2, matthias_2kf, ips-fflo, ips-nfl-u1}, where each of $\left \lbrace \mathbf K, \,k_{d-1} \right \rbrace $ has dimension unity and $k_d$ has dimension $1/2$. Dependence on $ e_k $ in the bosonic propagator will be generated via the susceptibility, which is obtained from the dynamics of the strong particle-hole fluctuations. Anticipating this, we have already added the extra term $\tilde a\,e_k   $, which will be generated in the RG process, and its has been dictated by the divergent term in the one-loop susceptibility calculated in the following subsection [see Eq.~\eqref{eqpi}]. In other words, we have simply included a term which will be generated via quantum corrections. This term has a mass dimension equal to unity, similar to the $k_d^2$ term, with a vanishing engineering dimension for $ \tilde a $. If we do not include this term, the loop integrations involving the bosonic propagator will turn out to be infrared-divergent, and these divergences will be the mere artifacts of our dropping the irrelevant terms in the minimal local effective action (if it contains only the $k_d^2$ term).
%In order to cure this, we need to resum a series of diagrams that dynamically generates non-trivial dependence along those frequency-momentum components. This is implemented by rearranging the perturbative expansion 
%such that the finite corrections from the one-loop boson self-energy are included at the `zero'-th order.
Finally, the engineering dimension of the fermion-boson coupling $e$ is equal to $x_e/2$ --- this observation has dictated us to introduce an explicit factor of a mass scale $\mu$ raised to the power $x_{e}/ 2$, chosen so as to ensure that $e$ is dimensionless, which is the usual procedure followed in QFT calculations.

The emergent \textit{sliding symmetry} in the Ising-nematic case \cite{max-isn,denis,ips-uv-ir1,ips-uv-ir2} forces the terms proportional to $\left[ \bar \Psi(k) \, k_{d-1}\, \Psi(k) \right ]$ and $\left [ \bar \Psi(k) \, k_{d}^2\, \Psi(k) \right ] $ in Eq.~\eqref{eqs1} to renormalize in the same way. In other words, the fermion propagator depends on $ k_{d-1}$ and $k_d^2 $
only through $\delta_k$ even after loop corrections. However, that is not the case here, and the renormalization process is not guaranteed to retain the sole dependence on $\delta_k$. In other words, it is possible that the nature of the renormalized terms may turn out to be such that it leads to a flattening of the Fermi surface at the hot-spots, as found in the RPA calculations of Ref.~\cite{metzner2}. Nevertheless, as we will see from our explicit calculations, the flattening does not show up in our one-loop level computations.

%%%%%%%%%%%%%%%%%%%%%%%%%%%%
\section{One-loop self-energies and dimensional regularization}
\label{secselfen}

The value of $x_e$ tells us that the coupling constant $e$ becomes marginal at the upper critical dimension $d_c=5/2$. In other words, $e$ is relevant for $d<5/2$ and irrelevant for $d >5/2$. Our aim is to access the interacting phase perturbatively in $d=5/2-\epsilon$, using $\epsilon $ as the perturbative parameter. In particular, this implies that at the end of our systematic $\epsilon$-expansion, we have to set $\epsilon=1/2$ for our original two-dimensional theory. Before embarking on deriving the RG flows, in this section, we compute the one-loop self-energies for the bosonic and fermionic degrees of freedom, which will feed into the equations necessary for determining the beta functions of the coupling constants $e$ and $\tilde a $.

The bare fermion and boson propagators for the action defined in Eq.~\eqref{eqs1} are given by
\begin{align}
G_{(0)} (k) & \equiv  \left\langle \Psi(k)\,  \bar{\Psi}(k) \right\rangle_0 
= \frac{1} {i}\,\frac{\bold{\Gamma} \cdot  {\mathbf{K}}  + \gamma_{d-1}\, \delta_k}
{ |\mathbf K|^2 + \delta_k^2} 
\,, \quad
D_{(0)}^+(k) \equiv  \left\langle \phi_+(k) \, \phi_-(-k) \right\rangle_0 
= \frac{1}{k_d^2 + \tilde a\, e_k } \,,\nn
D_{(0)}^- (k) & \equiv  \left\langle \phi_-(k)\, \phi_+(-k) \right\rangle_0 
= D_{(0)}^+(k) \,.
\end{align}

%%%%%%%%%%%%%%%%%%%
\begin{figure}[t]
\begin{center}
\subfigure[]{\includegraphics[scale=0.4]{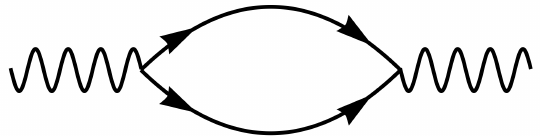} 
\label{fig:bos}}\hspace{1 cm}
\subfigure[]{\includegraphics[scale=0.5]{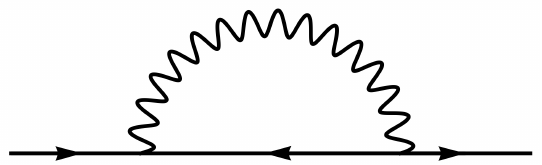} \label{fig:ferm}} \hspace{1 cm}
\subfigure[]{\includegraphics[scale=0.4]{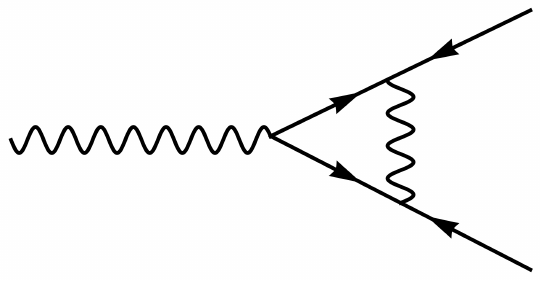} 
\includegraphics[scale=0.4]{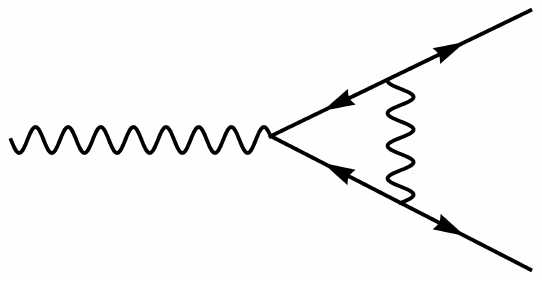} \label{fig:vert}}
\end{center}
\caption{The one-loop diagrams
for (a) the boson self-energy;
(b) the fermion self-energy; and (c) fermion-boson vertices.
Curves with arrows represent the bare fermion propagator $G_{(0)}$,
whereas the wiggly lines represent the dressed bosonic propagator $D_{(1)}$.
}
\label{fig:1loop}
\end{figure}

%%%%%%%%%%%%%%%%%%%%%%%%%%%%%%%%%%%%%
\subsection{One-loop boson self-energy}
\label{secbos}

The one-loop boson self-energy [cf. Fig.~\ref{fig:bos}] is defined by
\begin{align}
\Pi (k) =- \frac{e^2 \,\mu^{x_e} } {4} \times 2
\int_{q} \text{Tr} \left[ \gamma_0 \, G_{(0)}(q)\,\gamma_0\, G_{(0)}^T(k-q) \right]. 
\label{eqbos2}
\end{align}
Using the commutation relations between the gamma-matrices, and the identities
$\gamma_{d-1}^T=-\gamma_0\,\gamma_{d-1}\,\gamma_0$ and $\Gamma^T=-\gamma_0\,\Gamma\,\gamma_0$, 
we get
\begin{align}
\Pi (k) & =  e^2 \, \mu^{x_e}
\int_q 
\frac{  \vec Q \cdot (\vec Q - \vec K ) - \delta_{q} \,\delta_{k-q}
}
{ \left  ( \vec Q^2 + \delta_{q}^2 \right )
\left  [(\vec Q -\vec K)^2 + \delta_{k-q}^2 \right ]} \,.
\end{align}
%%%%%%%%%%%%%%%%%%%%%%%%%%%
Noting that
$\delta_{k-q} =  k_{d-1} +q_{d-1} +  (k_d-q_d)^2 $, we first shift $ {q}_{d-1} \rightarrow q_{d-1} -q_d^2$, and then use the Feynman parametrization to obtain
%%%%%%%%%%%%%%%%%%%
\begin{align}
\Pi (k) & =   e^2 \, \mu^{x_e} 
\int_q \int_0^1 dt\,
\frac{
|\mathbf Q|^2- t\, (1-t) \,|\mathbf K|^2
-\tilde e_{kq} \,q_{d-1} +  q_{d-1}^2}
{
 \left [
|\mathbf Q|^2  + t \,(1-t)\, |\mathbf K|^2
+ t\,\tilde e_{kq}^2 +  q_{d-1}^2 
-2 \,t \,\tilde e_{kq} \,q_{d-1} \right ]^2
}
\quad \big(\text{where } 
\tilde e_{kq} =  k_{d-1} +k_d^2-2 \,k_d \,q_d +2 \,q_d^2 \big )
\nn 
%%%%%%%%%%%%%%%%%
& =  e^2 \, \mu^{x_e} \int  d^{d-1} |\mathbf Q| \,dq_d
\int_0^1 dt\,
\frac{ |\mathbf Q|^d}
{2^{d} \,\pi^{\frac{d+1}{2}} \,
\Gamma \left(\frac{d-1}{2}\right)
\left[  |\mathbf Q|^2
 + t\,(1-t) \left(\tilde{e}_{kq}^2 + |\mathbf K|^2\right)\right]^{3/2}
} \nn 
%%%%%%%%%%%%%%%%%
& =   e^2 \, \mu^{x_e} 
\int  dq_d \,
\frac{2^{ 1-2\, d} \csc \big(\frac{ d\,\pi} {2}\big) 
\left(\tilde{e}_{kq}^2 + |\mathbf K|^2\right)^{\frac{d-2}{2}
} }
{\pi^{\frac{d-1}{2}} \,\Gamma \left(\frac{d-1}{2}\right)}\,.
\end{align}
%%%%%%%%%%%%%%
Changing variables to $ u =\sqrt 2\,q_d -k_d /\sqrt 2 $, with the Jacobian factor $1/\sqrt 2$,
we get
\begin{align}
\Pi (k) & =  e^2 \, \mu^{x_e} 
\int_0^\infty du\,
\frac{2^{\frac{3}{2}-2\, d} \csc \big(\frac{ d\,\pi }{2}\big) 
\left[ \left(u^2 +  e_k \right )^2
+ |\mathbf K|^2\right]^{\frac{d-2}{2}}
}
{\pi^{\frac{d-1}{2}} \,\Gamma \left(\frac{d-1}{2}\right)}\,.
\end{align}
Since the bare susceptibility at zero temperature diverges at the
nesting vector $\mathbf Q$, the well-defined self-energy is given by subtracting off the singular contribution for this momentum, which translates to
\begin{align}
\label{eqtildepi}
& \tilde \Pi (k)  =  \Pi (k)-  \Pi (0)
= \frac{ 2^{\frac{3}{2}-2 d} \csc \big(\frac{ d\,\pi }{2}\big) \,
e^2 \,\mu^{x_e}}
{\pi^{\frac{d-1}{2}} \,\Gamma \left(\frac{d-1}{2}\right)
}
\,I_{\Pi}(k,d)\,,
\end{align}
where (after changing variables as $z = u^2 + e_k $)
\begin{align}
I_{\Pi} (k,d) & \equiv \int_{e_k}^\infty dz\,
\frac{
\left ( z-{  e_k} \right )^{2-d}
-
\left[ z^2+ |\mathbf K|^2\right]^{\frac{2-d}{2}}
}
{ 2\,\sqrt{z-{  e_k}}
\, \left[ z^2+ |\mathbf K|^2\right]^{\frac{2-d}{2}}
\, \left ( z-{  e_k} \right )^{2-d}
}\nn
%%%%%%%%%%%%%%%%%%%%%%%%%%%%%%%%%%%
& =\begin{cases}
\frac{ \Gamma (d-1) \,
 \left(-e_k\right)^{d-\frac{3}{2}}}
{ 2 } 
\left [ \frac{\Gamma \left(\frac{3-d}{2}\right) \, 
\, _2F_1\left(\frac{ 3-2 \,d}{4} ,
\frac{5-2 \,d}{4} ;\frac{3-d}{2};
-\frac{|\mathbf K|^2}{ { e_k}^2}\right)}{\sqrt{\pi }}
+ \frac{_2 {F}_1\big (\frac{1}{2},1;d;1 \big) } {\Gamma(d)}
\right]  & \\
%%%%%%%%
+ \, \frac{ |\mathbf K|^d \,\,
 _3F_2\big (\frac{3}{4},1,\frac{5}{4};\frac{3}{2},\frac{d}{2}+1;
 -{|\mathbf K|^2}/ { { e_k}^2}\big )}
{  4 \,d\, \left(- { e_k} \right)^{3/2}}
+
\frac{\pi^{3/2} \,
|\mathbf K|^{d-1} \sec \left(\frac{ d\,\pi} {2}\right)  
\, \,  _2 {F}_1 \big(\frac{1}{4},\frac{3}{4};\frac{d+1}{2};
-{|\mathbf K|^2} /{ e_k^2}\big)}
{ 4\, \sqrt{- { e_k}  } \,\,\Gamma \left(\frac{2-d}{2}\right)
\, \Gamma \left (\frac{d+1}{2} \right )
} & \\
%%%%%%%%%%%%%  
+ \, |\mathbf K|^{d-2} \,\sqrt{-e_k} \, \,
_3F_2\left(\frac{1}{2},1,1-\frac{d}{2};\frac{3}{4},\frac{5}{4};-\frac{e_k^2} {|\mathbf K|^2}\right)
+ \frac{ \left(-e_k\right)^{d-\frac{3}{2}} }
{3-2\, d}
& \text{ for }  { e_k}<0
%%%%%%%%%%%%%%%%%%%%%%
%%%%%%%%%%%%%%%%%%%%%%%%%%%%
\\ & \\ & \\
\frac{\sqrt{\pi } \,  e_k^{ d - \frac{3}{2}} 
\, \Gamma \left(\frac{3-d}{2}\right) \, \,
_2F_1 \big (\frac{3-2 \,d}{4} ,\frac{5-2 \,d}{4} ;\frac{3-d}{2};-
 |\mathbf K|^2 / { {e_k}^2} \big )} 
 {2\, \Gamma (2-d)}
& \text{ for }   e_k  > 0\,.
\end{cases} 
\end{align} 
Note that the zeroes coming from $1/\Gamma\big (\frac{2-d} {2} \big)$ and $1/\Gamma(2-d)$ at $d=2$ are cancelled by the factor $ \csc \big(\frac{ d\,\pi }{2}\big)$ present in Eq.~\eqref{eqtildepi}.

The leading-order terms, obtained in the limit $|\mathbf K|^2 / {  e_k^2} \ll 1 $, are found to be:
\begin{align}
I_{\Pi} (k,d) & =\begin{cases}
\frac{ \Gamma (d-1) \,(-  e_k  )^{ d -\frac{3}{2}  }
}
{2 }
\left [  
\frac{ _2\tilde{F}_1 \big (\frac{1}{2},1;d;1 \big )}  {\Gamma(d)}
+ \frac{\Gamma \left(\frac{3}{2} - d \right)}
{\sqrt{\pi }}
+ \frac{\sqrt{\pi }}{\Gamma \left(d-\frac{1}{2}\right)}
\right ]
+\frac{(-  e_k  )^{ d -\frac{3}{2}  }} 
{3-2 \,d}
+
\frac{\pi^{3/2} \,  |\mathbf K|^{d-1} 
\sec \left(\frac{ d\,\pi} {2}\right )}
{ 2 \,\sqrt{-  e_k  } \,
\,\Gamma \left(\frac{2-d}{2} \right)  \,\Gamma \left(\frac{d+1}{2}\right)
}
+ \mathcal{O}\left(\frac{|\mathbf K|^d}  {(- e_k )^{3/2}} \right)
& \text{ for }   e_k  < 0 \\
& \\
%%%%%%%%%%%%%%
%%%%%%%%%%%%%%%%%%%%%%
\frac{\sqrt{\pi }  \,
\Gamma \left(\frac{3}{2}-d\right)
\,e_k^{ d -\frac{3}{2} }} 
{2 \,\Gamma (2-d) }
+
\frac{
\sqrt{\pi } \,\Gamma \left(\frac{7}{2}-d\right)
\, |\mathbf K|^2 }
{ 4 \,(d-3) \,\Gamma (2-d)\,  e_k ^{\frac{7-2\,d} {2}}
} 
+ \mathcal{O}\left( \frac{|\mathbf K|^4}  {  e_k^{11/2-d}}  \right)
& \text{ for }   e_k  > 0\,.
\end{cases} 
\end{align} 
%%%%%%%%%%%%%%%%%%%%%%%
The factors $\Gamma \big ( 3/2-d \big )$ and $1/(3-2d)$ have poles at $ d= 3/2$, which shows that the terms containing them term have (1) logarithmic divergence at $d = 3/2$, and (2) linear divergence at $d= 5/2 $, when we translate the divergences in the language of the Wilsonian cutoff $\Lambda \sim \mu $. We need to treat this result carefully by remembering that, in the dimensional regularization procedure, UV divergences of all degrees show up as the poles of $\Gamma$-functions. The degree of divergence can be understood by using an explicit UV cut-off in the language of the Wilsonian RG, which is denoted here by $\Lambda$. Although these terms will play an important role for analyzing UV-stable fixed points, they should be discarded here because we are considering the RG flows in the IR. In fact, these terms represent IR-irrelevant operators for the theory in $d= 5/2 -\epsilon$ spatial dimensions.
% reflected by the that they have a scaling dimension equal to $(1-\epsilon)$ at $ d=5/2-\epsilon$. 
Some more comments are in order. The situation above is similar to having a $\phi^6$-term in a $\phi^4$ scalar field theory in $(3+1)$-dimensions. A simple power counting of momenta arising in loop diagrams shows that adding a $\phi^6$-interaction vertex to a free scalar field theory gives an upper critical dimension of $3$, while addition of a $\phi^4$-vertex has the upper critical dimension value of $4$. Therefore, adding a $\phi^6$-vertex to the $\phi^4$ theory in four spacetime dimensions makes the theory nonrenormalizable.
%The above statements can also be quantified in an alternate way. Let us expand the expression in $\tilde \epsilon $, where $d= 3/2- \tilde \epsilon $. Then we have
%\begin{align}
%\Sigma_2 \Big \vert_{d=3/2-\tilde \epsilon } 
%=  \frac{ 
%e^2\, \left[ 
% - e_k
%+ \left( \frac{  6 +  \tilde{a} }
%{  2 - \tilde{a} } \right)  \frac{k_d^2} {2}
% \right ]
%}
%{  \sqrt 2 \, \pi^{5/4}
%\left( 2 -\tilde{a}\right)  \, \tilde \epsilon }
%%%%
%\left [  \frac{\mu}
%{ - e_k
%+ \left( \frac{  6 +  \tilde{a} }
%{  2 - \tilde{a} } \right)  \frac{k_d^2} {2}}
%  \right ]^{1 + \tilde \epsilon} + \order { \tilde \epsilon^0} ,
%\end{align}
%%%%%%%%%%%%%%%%%%%%%%%%%%%%
%which indicates that the term has a linear UV divergence at $d = 5/2$.
%Another crucial observation is that $I_3 $ vanishes identically at $d=2$. All these observations together dictate that there will be no contribution to the RG flows in the IR from these term and, hence, should not be included in the counterterms. 

Finally, expanding in $\epsilon$ for $d=5/2-\epsilon $, after discarding the terms containing poles at $ d=3/2 $, we get
\begin{align}
 \left[ \mu^{x_e} \,
\frac{ 2^{\frac{3}{2}-2\, d} 
\csc \big(\frac{ d\,\pi }{2}\big) }
{\pi^{\frac{d-1} {2}} \,\Gamma \left(\frac{d-1}{2}\right)}
\times I_{\Pi} (k,d) \right ] \Bigg \vert_{d= \frac{5} {2} -\epsilon} 
%%%%%%%%%
& =\begin{cases}
- \,\frac{ \pi^{3/4} \,
\frac{|\mathbf K|^{\frac{3}{2} }}  
{ \sqrt{|  e_k |} } 
\,  \,\left( \frac{\mu} { |\mathbf K|} \right)^\epsilon
}
{ 32 \,\sqrt{2} \, \Gamma^2 (3/4)  
\,\Gamma \left(7/4\right) }
%%%
 + \mathcal{O}\left(\epsilon \right )
& \text{ for }   e_k  < 0 \\ & \\
%%%%%%%%%%%%%%%%%%
-\,\frac{ \frac{|\mathbf K|^2 } {  e_k }
\, \, \left( \frac{\mu } {  e_k } \right)^\epsilon
}
{ 32 \,\pi^{3/4} \,\Gamma \left( 3/4\right) 
}
+ \mathcal{O}\left(\epsilon \right )
& \text{ for }   e_k  > 0
\end{cases} .
\end{align} 
%%%%%%%%%%%%%%%%%%%%%%
We conclude that the leading-order term for the self-energy correction in Eq.~\eqref{eqtildepi} is given by
%%%%%%%%%%%%%%%%%%%%%
\begin{align}
\label{eqpi}
\tilde \Pi(k) = 
- \,\frac{  e^2 \, \mu^{x_e} \, b  \, |\mathbf K|^{d-1}  }   
 { \sqrt { | e_k |} }
\,\Theta(- e_k )  \,,
\text{ where }
%%%%%%%%%%%%%%
b & = \frac{ \pi^{3/4} }
{ 32 \,\sqrt{2} \, \,\Gamma^2 (3/4)  
\,\Gamma \left(7/4\right) } \,,
\end{align}
%%%%%%%%%%%%%%%%%%%%
in the limit $|\mathbf K|^2 / {  e_k^2} \ll 1 $.

We note that since the bare boson propagators $D_{(0)}^\pm (k) $ are independent of $\mathbf K$, the loop
integrations involving them are ill-defined, unless one resums a series of diagrams that provides a nontrivial dispersion along these frequency and momentum components. Hence, in all loop calculations involving the boson propagators, we include the finite correction [denoted by $ \tilde \Pi(k) $] from the one-loop self-energy, which is proportional to $ |\mathbf K|^{d-1} /  \sqrt { | e_k |} $. Therefore, both for the $ \phi_+(k)$ and $\phi_- (k) $ bosonic fields, we use the dressed propagator
\begin{align}
\label{eqbos1}
D_{(1)} (k) = \frac{1}
{ \left[ D_{(0)}^+(k) \right]^{-1}- \, \tilde \Pi (k) }
=\frac{1}
{k_d^2 +  \frac{ b\, e^2\,\mu^{x_e} \,|\mathbf K|^{d-1} \,\Theta(-e_k)
} 
{\sqrt{| e_k|}} }\,,
\end{align}
%%%%%%%%%%%%%%%%%%%%%%%%%%%%%%%
which is equivalent to rearranging the perturbative loop-expansions such that the one-loop finite part of the boson self-energy, dependent on $ \mathbf K $, is included at the zeroth order. We would like to emphasize that $ \tilde \Pi(k) $ is the so-called {\textit{Landau-damped term}} which leads to the signature $\text{sgn} (k_0) |k_0|^{2/3} $-dependence of the fermion self-energy, characterizing the non-Fermi liquid behaviour in various quantum critical systems \cite{max-isn, max-sdw, denis, ips-uv-ir1, ips-uv-ir2, matthias_2kf, ips-fflo, ips-nfl-u1}. The Landau-damped part also plays the most significant role in inducing unconventional superconductivity in this kind of non-Fermi liquid systems \cite{ips2, ips3, Max_sc, ips-sc, ips-sc_err}.

%%%%%%%%%%%%%%%%%%%%%%%%%%%%%%%%%%%%%
\subsection{One-loop fermion self-energy}
\label{secferm}

The fermion self-energy [cf. Fig.~\ref{fig:ferm}] is given by the integral
\begin{align}
\label{eqfs0}
\Sigma(k) & =  e^2 \,\mu^{x_e}
\int_{q} \,\gamma_0\, G_{(0)}^T(q-k) \,\gamma_0\, D_{(1)}(q)  
 % = - e^2 \,\mu^{x_e} \int_{q} \, G_{(0)}(q-k) \,  D_{(1)}(q) 
%%%%%%%%
%=
%i \, e^2 \,\mu^{x_e} 
%\int_q \frac{- \mathbf{\Gamma} \cdot (\mathbf Q- {\mathbf{K}} ) 
%+ \gamma_{d-1} \, \delta_{q-k}}{(\mathbf Q- {\mathbf{K}} )^2 
%+ \delta^2_{q-k}} \, D_{(1)}(q) \nn
%%%%%%%%%%%%
 = i\,\Sigma_1 (k) \, \mathbf{\Gamma} \cdot \mathbf K
+ i\,\Sigma_2 (k) \,\gamma_{d-1} \,,
\end{align}
where
\begin{align}
\Sigma_1 (k) = - \frac{e^2 \,\mu^{x_e} } { |\mathbf K|^2 }
\int_q \frac{ \mathbf{K} \cdot (\mathbf Q- {\mathbf{K}} ) 
}
{(\mathbf Q- {\mathbf{K}} )^2 
+ \delta^2_{q-k}} \, D_{(1)}(q)
\end{align}
and
\begin{align}
\label{eqfs2}
\Sigma_2 (k) =  e^2 \,\mu^{x_e} 
\int_q \frac{ \delta_{q-k} }
{(\mathbf Q- {\mathbf{K}} )^2 + \delta^2_{q-k}} \, D_{(1)}(q)\,.
\end{align}
The steps to compute these two parts have been explained in the next two subsections, which can be skipped if the reader is not interested in the tedious intermediate steps. For their benefit, we state here the final result. Setting $d = d_c-\epsilon$, we get the singular part to be
\begin{align}
\Sigma(k) & =- 
 \frac{ e^{4/3} \, \, {\mathcal U}_1 \, } 
{ \left (2- \tilde a  \right )^{2/3}
\,\epsilon} 
\, i\left( \mathbf{\Gamma} \cdot \mathbf K \right)
+\order{\epsilon^0} ,\quad
%%%%%%%%%%%%%%%%
{\mathcal U}_1 = \frac{\sqrt{2} \,\,
 \Gamma \big (\frac{5}{4}\big)}
 {3\, \sqrt 3 \, \pi^{7/4} \, b^{1/3}  }\,,
\label{eqferm1}
\end{align}
where the divergence is parametrized by a pole at $  \epsilon =0 $.

%%%%%%%%%%%%%%%%%%%%%%%%%%%%%%%%%%%%
\subsubsection{Computation of $\mathbf \Gamma$-dependent part}
%%%%%%%%%%%%%%%%%%%%%%%%%%%%%%%%%%%%%

The leading order dependence of $\Sigma_1 (k)$ on $\mathbf K$ can be extracted by setting the external momentum components $k_d$ and $k_{d-1}$ to zero. Hence, we will evaluate
\begin{align}
\Sigma_1 (\mathbf K, 0,0)  =  \frac{ e^2 \,\mu^{x_e} } 
{ |\mathbf K|^2 }
\int_q \frac{ \mathbf{K} \cdot (\mathbf K - {\mathbf{Q}} ) 
}
{(\mathbf Q- {\mathbf{K}} )^2 
+  \delta_q^2 }  \times
\frac{1}
{ q_d^2 +  e^2\,\mu^{x_e} \, b\, |\mathbf Q |^{d-1} \,\Theta(- e_q)
 /\sqrt{| e_q|}  } \,.
\end{align}
%%%%%%%%%%%%%%%%%%%555
Changing the description to $ q_d $ and $  e_q$ as integration variables, and dividing into the parts $  e_q< 0 $ and $  e_q> 0$ as
$ { \Sigma_1 (\mathbf K, 0,0)   
} = I_1 + I_2 $, we have
%%%%%%%%%%%%%%%%%%%% 
\begin{align}
I_1 & = \frac{ e^2 \,\mu^{x_e} } 
{ |\mathbf K|^2 }
\int_{ e_q<0} \frac{d^{d-1} \mathbf{Q}\, dq_d \,d e_q
}
{ (2\,\pi)^{d+1} }  \,
 \frac{ - \mathbf{K} \cdot (\mathbf Q - \mathbf K ) }
{(\mathbf Q- {\mathbf{K}} )^2 
+ \left(  e_q+ q_d^2/2  \right)^2
}  \,
\frac{1}
{ q_d^2  + e^2 \,\mu^{x_e} \, b \,
|\mathbf Q |^{d-1}  /\sqrt{| e_q|}  } \nn
%%%%%%%%%%%%%%%%%%%
 & =  \frac{ e^2 \,\mu^{x_e} } 
{ |\mathbf K|^2 }
 \int_0^\infty \frac{ du} {\sqrt { u/2} }
 \int_0^\infty d e_q
  \int_{-\infty}^{\infty}
  \frac{d^{d-1} \mathbf{Q}
}
{  (2\,\pi)^{d+1} }\,
 \frac{ \mathbf K^2 -\mathbf K \cdot \mathbf Q
}
{(\mathbf Q- {\mathbf{K}} )^2 
+   \left( u- e_q  \right)^2
}  \,
\frac{1}
{ 2\, u  + e^2\,\mu^{x_e}\, b \,|\mathbf Q |^{d-1} /\sqrt{ e_q} 
}  \,\, \left [\text{where } 2\,u = {q_d^2} 
\right ],
\end{align} 
and
%%%%%%%%%%%%%%%%%%%% 
\begin{align}
I_2 & =  \frac{ e^2 \,\mu^{x_e} } 
{ |\mathbf K|^2 }
\int_{ e_q>0} \frac{d^{d-1} \mathbf{Q}\, dq_d \,d e_q
}
{ (2\,\pi)^{d+1} }  \,
 \frac{ - \mathbf{K} \cdot (\mathbf Q - \mathbf K ) 
}
{(\mathbf Q- {\mathbf{K}} )^2 
+ \left(  e_q+ q_d^2/2  \right)^2
}  \,
\frac{1}
{ q_d^2 }
%%%%%%%%%%%%%
\nn & = 
\frac{ e^2 \,\mu^{x_e} } 
{ |\mathbf K|^2 }
\int_{ e_q>0} \frac{d^{d-1} \mathbf{Q}\, dq_d \,d e_q
}
{ (2\,\pi)^{d+1} }  \,
 \frac{ - \mathbf{K} \cdot \mathbf Q 
}
{ \mathbf Q^2  + \left(  e_q+ q_d^2/2  \right)^2
}  \,
\frac{1}{ q_d^2 } = 0 \,.
\end{align}

The integral $I_1$ cannot be evaluated exactly and we need to make some reasonable approximations to extract the leading order corrections. We note that the first factor of the integrand tells us that the dominant contribution is concentrated around the region $ |\mathbf Q | \sim |\mathbf K |$ and $   u \sim e_q  $. As for the second factor, the dominant contribution comes from $ e_q  \sim |\mathbf Q |^{2\,(d-1)/3} \sim  |\mathbf K |^{2\,(d-1)/3}$. Since $|\mathbf K |^{2\,(d-1)/3} \gg |\mathbf K |$ for small $|\mathbf K|$ close to zero and $2\,(d-1)/3 <1 $, we can substitute $u \sim  e_q$ in the $\sqrt u $ factor in the overall denominator and the $2\,u$ term in the denominator of the second factor, and extend the lower limit of the integral over $u$ to $-\infty$, leading to
%%%%%%%%%%%%%%%%%%%%%%%
\begin{align}
I_1 & \simeq \frac{ e^2 \,\mu^{x_e} } 
{ |\mathbf K|^2 }
\int_{-\infty}^\infty 
\frac{d^{d-1} \mathbf{Q}\, du} {(2\,\pi)^{d+1}}
  \int_{e_q>0} \frac{  d e_q}
{   \sqrt{ e_q/2 } }\,
 \frac{ \mathbf{K} \cdot (\mathbf K - {\mathbf{Q}} ) 
}
{(\mathbf Q- {\mathbf{K}} )^2 +  u^2
}  \,
\frac{1}
{ 2\,  e_q
+ e^2\, \mu^{x_e}\, b\,|\mathbf Q |^{d-1} /\sqrt{ e_q } }
\quad \left [\text{shifting } u \rightarrow u  + e_q
\right ] \nn
%%%%%%%%%%%%%%%%%%%%%%%
& =  \frac{ e^2 \,\mu^{x_e} } 
{ |\mathbf K|^2 }
\int_{-\infty}^\infty 
\frac{d^{d-1} \mathbf{Q}\, du} {(2\,\pi)^{d+1}}
  \int_{e_q>0} d e_q \,
 \frac{ \mathbf{K} \cdot (\mathbf K - {\mathbf{Q}} ) 
}
{(\mathbf Q- {\mathbf{K}} )^2 + u^2
}  \,
\frac{ \sqrt 2
}
{ 2\,   e_q^{3/2}
+  e^2\, \mu^{x_e}\, b \,|\mathbf Q |^{d-1} } \nn
%%%%%%%%%%%%%%%%%%%%%%%%%%%%%%%%%
& = -\frac{
e^{4/3} \, \Gamma \big (\frac{5-2 \,d}{6} \big ) \,
\Gamma \big (\frac{d}{2}\big ) \,
\Gamma \big (\frac{d+2}{6} \big ) 
} 
{ 2^{\frac{4 \,d-1} {6} } \pi^{\frac{d+1}{2}} \times 
3 \, \sqrt{3} \times 2^{2/3}
\, b^{1/3} \, \Gamma \big (\frac{5 \,d-2}{6} \big )}
 \left(  \frac{\mu} { |\mathbf K| } \right)^{\frac{2\, x_e} {3} }\,.
\end{align}
%%%%%%%%%%%
The integral blows up at $d = 5/2$, which thus gives us the value of the upper critical dimension $ d_c $.\
The fermion-boson coupling $e$ is irrelevant for $ d> d_c$, relevant
for $d < d_c$, and marginal for $d = d_c$. This allows us to access the strongly interacting non-Fermi liquid state perturbatively, in a controlled approximation, using $ d = 5/2 -\epsilon $, where $\epsilon$ serves as the perturbative/small parameter. In our dimensional regularization scheme, the divergence appears as $\sim \epsilon^{-1} $, with the $ \Gamma \big (\frac{5-2 \,d}{6} \big )$ factor having a pole at $d =d_c$.
We also note that this term produces the behaviour of the fermion self-energy as $ \sim \text{sgn}(k_0) \,|k_0|^{ 2/3} $ at $d=2$, which matches with the uncontrolled RPA result \cite{metzner1, metzner2}. We would like to point out that the correct $k_0$-dependence of $\Sigma$ could be captured only because we have included the crucial Landau damping term in the dressed bosonic propagator $D_{(1)}$. The $ |k_{0}|^{ 2/3} $-scaling was missed in Ref.~\cite{matthias_2kf}
due to the non-inclusion of $\Pi_{\rm LD}$.

%%%%%%%%%%%%%%%%%%%%%%%%%%%%%%%%%%%%
\subsubsection{Computation of $\gamma_{d-1} $-dependent part}
%%%%%%%%%%%%%%%%%%%%%%%%%%%%%%%%%%%%%

The leading order dependence of $\Sigma_2 (k)$ on $ k_d$ and $k_{d-1} $ can be extracted by setting $\mathbf K= 0 $. Hence, we will evaluate
\begin{align}
\label{eqfs20}
\Sigma_2 (\mathbf 0, k_d , k_{d-1})  & =  e^2 \,\mu^{x_e} e^2 \,\mu^{x_e} 
\left[ I_3 (k_d , k_{d-1}) + I_4 (k_d , k_{d-1}) \right ] ,
\end{align}
%%%%%%%%%%%%%%%%%%5
where
$ \delta_{q-k}  =  q_{d-1} - k_{d-1} +k_d^2 + q_d^2  - 2 \,k_q\, q_d $
and
\begin{align}
I_3 (k_d , k_{d-1})= \int_{q, \, e_q <0} \frac{ \delta_{q-k} }
{ |\mathbf Q|^2 + \delta^2_{q-k}}   \times \frac{1} 
{ q_d^2 + e^2 \, \mu^{x_e} \, b \,|\mathbf Q|^{d-1}  / \sqrt{| e_q|} } 
\text{ and }
I_4 (k_d , k_{d-1})= \int_{q, \, e_q > 0} \frac{ \delta_{q-k} }
{ |\mathbf Q|^2 + \delta^2_{q-k}}  \times \frac{1} {  q_d^2 } \,.
\end{align}

To simply the calculations a bit, we set $k_d $ to zero. Therefore,
\begin{align}
I_3 ( 0 , k_{d-1}) & = \int_{q, \, e_q > 0} \frac{  q_d^2 /2  -e_q  - k_{d-1}   }
{ |\mathbf Q|^2 + \left( q_d^2 /2  - e_q - k_{d-1}  \right)^2 }  \times \frac{1} 
{  q_d^2  +  e^2 \, \mu^{x_e} \, b \,|\mathbf Q|^{d-1}  /\sqrt{e_q} } \nn
%%%%%%%%%%%%%%%%%%%%%%%%%%
& = \int_0^\infty \frac{ du} {\sqrt { u/2} }
 \int_0^\infty d e_q
  \int_{-\infty}^{\infty} \frac{d^{d-1} \mathbf{Q}}
{  (2\,\pi)^{d+1} }\, \frac{ u -e_q  - k_{d-1} }
{ |\mathbf Q|^2 
+   \left(  u -e_q  - k_{d-1}  \right)^2 }  \,
\frac{1}   { 2\, u  + e^2\,\mu^{x_e}\, b \,|\mathbf Q |^{d-1} /\sqrt{ e_q} }  
\; \left (\text{where } 2\,u = {q_d^2} \right ) .
\end{align}
%%%%%%%%%%%%%%%%%%%%%%%%%%%%%%%%%%%%%%%%%%
We observe that the first factor of the integrand forces the dominant contribution to be concentrated around the region $ |\mathbf Q | \sim 0 $ and $   u \sim e_q +  k_{d-1}  $. As for the second factor, the dominant contribution comes from $ e_q  \sim |\mathbf Q |^{2\,(d-1)/3}$. Hence, we can substitute $u \sim  e_q  + k_{d-1}$ in the $\sqrt u $ factor (in the overall denominator) and the $2\,u$ term (in the denominator of the second factor), and extend the lower limit of the integral over $u$ to $-\infty$. This leads to
%%%%%%%%%%%%%%%%%%%%%%%
\begin{align}
I_3 ( 0 , k_{d-1}) \simeq \int_0^\infty \frac{ du} {\sqrt {e_q +  k_{d-1}} }
 \int_0^\infty d e_q
  \int_{-\infty}^{\infty} \frac{d^{d-1} \mathbf{Q}}
{  (2\,\pi)^{d+1} }\, \frac{ u -e_q  - k_{d-1} }
{ |\mathbf Q|^2 
+   \left(  u -e_q  - k_{d-1}  \right)^2
}  \, \frac{ \sqrt 2}
{  2 \, e_q + 2\, k_{d-1} + e^2\,\mu^{x_e}\, b \,|\mathbf Q |^{d-1} /\sqrt{ e_q} }  = 0 \,.
\end{align}
Hence, to leading order, the $I_3$ integral vanishes, and there remains no $b$-dependent term.
Next, we have
%%%%%%%%%%%%%%%%%%%%%%%
\begin{align}
I_4 ( 0 , k_{d-1})= \int_{q, \, e_q > 0} \frac{  q_d^2 /2  + e_q  - k_{d-1}   }
{ |\mathbf Q|^2 + \left( q_d^2 /2 + e_q - k_{d-1}  \right)^2 }  \times \frac{1} {  q_d^2 }
=
\int_0^\infty \frac{ du} {\sqrt { u/2} }
 \int_{-\infty}^0 d e_q \int_{-\infty}^{\infty} \frac{d^{d-1} \mathbf{Q}}
{  (2\,\pi)^{d+1} }\, \frac{ u + e_q  - k_{d-1} }
{ |\mathbf Q|^2 
+   \left(  u + e_q  - k_{d-1}  \right)^2 }  \,
\frac{1}   { 2\, u}  \,.
\end{align}
We apply a similar argument as before, namely the first factor of the integrand forces the dominant contribution to be concentrated around the region $ |\mathbf Q | \sim 0 $ and $   u \sim   k_{d-1} - e_q $. Substituting $u \sim  k_{d-1} - e_q $ in the $\sqrt u $ factor (in the overall denominator) and the $2\,u$ term (in the denominator of the second factor), and extending the lower limit of the integral over $u$ to $-\infty$, we get
\begin{align}
I_4 ( 0 , k_{d-1}) = 0\,.
\end{align}
Consequently, the flattening of the Fermi surface at the hot-spots, found in the RPA calculations~\cite{metzner2}, does not show up in our one-loop results.

%%%%%%%%%%%%%%%%%%%%%%%%%%%%%%%%%%%%%
\subsection{One-loop vertex correction}
\label{secvert}

The one-loop vertex diagrams [cf. Fig.~\ref{fig:vert}] yield non-divergent corrections and, hence, do not contribute to renormalization and RG flows.

%%%%%%%%%%%%%%%%%%%%%%%%%%%%%%%%%%%%%%%%%5
\section{Renormalization Group flows under minimal subtraction scheme}
\label{secrg}

Eq.~\eqref{eqs1} is supposed to be the \textit{physical action}, defined at an energy scale $ \mu \sim \Lambda $, consisting of the fundamental Lagrangian with non-divergent quantities. However, we have seen that the loop integrals lead to divergent terms and, in order to cure it, we employ the renormalization procedure, using dimensional regularization as the regularization method. In our dimensional regularization formalism, the UV-divergent terms are the ones arising in the $\epsilon \rightarrow 0$ limit. We use the minimal subtraction ($ {\rm MS}$) renormalization scheme to control the UV divergences \cite{thooft, weinberg}, which involves cancelling the divergent parts of the loop-contributions via adding appropriate counterterms. More precisely, we adopt the modified minimal subtraction ($\overline{\rm MS}$) scheme where, in addition to the divergent term, we absorb the universal term proportional to $\epsilon^ 0$ (that always accompanies the term with the $ 1/\epsilon $ pole) into the corresponding counterterm.

The action, consisting of the counterterms to absorb the singular terms, takes the form:
\begin{align}
\label{actcount}
S_{CT}  = &   \int_k \bar \Psi (k)
\, i \,\Bigl[ 
 A_{1} \,{\vec \Gamma} \cdot { \vec K} 
+   \gamma_{d-1} \left( A_2 \,e_k
+ A_3 \,\frac{ k_d^2 } {2} \right )   \Bigr] \Psi (k) 
%%%%%%%%%%%%%%%%%%%%%%%%%%%  
 + \int_k 
 A_4 \, k_d^2  \,\phi_+ (k) \, \phi_-(-k)  \nn & \,
- \frac{ i\, e \, \mu^{x_e/2} } {2} 
\int_{k, \,q} A_6 \,
\Big[ \,\phi_+(q) \,
 \bar{\Psi} (k+q) \, \gamma_0 \, \bar{\Psi}^T(-k) 
- \phi_-(-q) \,\Psi^T(q-k) \,\gamma_0 \,\Psi(k) \Big] \,.
\end{align}
%%%%%%%%%%%%%%%%%%%%%%%%%%%%%%
The counterterm-factors are given by the power series
\begin{align}
A_{\zeta} = 
\sum_{ n=1}^\infty \frac{Z^{(n)}_{ \zeta}}
{\epsilon^n}  \text{  with }  
\zeta \in [1, 5]\,,
\end{align}
such that they cancel the divergent $1/\epsilon^n$ contributions from the Feynman diagrams.
Due to the $(d-1)$-dimensional rotational invariance in the space perpendicular to the Fermi surface, each term in $ {\vec \Gamma} \cdot {\vec K}$ is renormalized in the same way.

Subtracting $S_{CT}$ from the so-called \textit{bare} action $S_{\text{bare}}$, we obtain the renormalized action, which is the \textit{physical} effective action of the theory, re-written in terms of non-divergent quantum parameters. While the bare parameters can be divergent, the physical observables are the renormalized coupling constants, which are determined by the RG equations.  The RG flows describe the evolution of the couplings as functions of the floating energy scale $\mu \,e^{- l} $ (i.e., with respect to an increasing logarithmic length scale $l$).
To achieve this objective, we first define the bare (or fundamental) action
%%%%%%%%%%%%%%%%%%%%%%%%
\begin{align}
\label{actren}
S_{\text{bare}}  = &   \int_{k^B} \bar{\Psi}^B (k^B)
\,\, i \left[ 
\,{\vec \Gamma} \cdot { \vec K^B} 
+   \gamma_{d-1} \left \lbrace  e_k^B + 
\frac{ \left( k^B\right)^2 }{2} \right \rbrace  \right ] \Psi^B (k^B)
%%%%%%%%%%%%%%%%%%%%%%%%%%%  
+ \int_{k^B}  \left (k^B_d \right)^2 \,
\phi_+^B (k^B) \,\, \phi_-^B (-k^B) 
 \nn & \,
 %%%
- \frac{ i\, e^B } {2} 
\int_{k^B,\, \, q^B}
\Big[ \,\phi^B_+(q^B) \,
 \bar \Psi^B  (k^B+q^B) \, \gamma_0 
 \left( {\bar \Psi}^B (-k^B) \right )^T  
- \phi^B_-(-q^B) \left( {\Psi^B} (q^B-k^B) \right)^T  \gamma_0 \,
\Psi^B (k^B) \Big] \,,
\end{align}
%%%%%%%%%%%%%
consisting of the \textit{bare quantities},
where the superscript ``$B$'' has been used to denote the bare fields, couplings, frequency, and momenta. 
We now relate the bare quantities to the so-called renormalized quantities
(without the superscript ``$B$'') via the multiplicative $Z_\zeta $-factors such that
\begin{align}
S_{\text{bare}}  = & S + S_{CT}\,, 
\quad Z_{\zeta}  =  1 + A_{\zeta}\,,
\end{align}
%%%%%%%%%%%%%%%%%%%
\begin{align}
&  {\vec K}^B =   
\frac{Z_1} {Z_3} \, {\vec K} \, , \quad
e_k^B = \frac{Z_2} {Z_3} \,  e_{k}  \, , 
\quad  k^B_d  =  k_d \,, \quad
\Psi^B(k^B)  =   Z_{\Psi}^{1/2}\, \Psi(k)\,, 
\quad \phi_{\pm}^B(k^B) =  Z_{\phi}^{1/2}\, \phi_{\pm}\,,
%%%%%%%%%%
\end{align}
and
\begin{align}
& Z_{\Psi}  =  Z_1 \left(\frac{Z_1}{Z_3}\right)^{-d} 
\left(\frac{Z_2}{Z_3}\right)^{-1}  ,\quad
Z_{\phi}  =  Z_4 
\left(\frac{Z_1}{Z_3}\right)^{1-d} 
\left(\frac{Z_2}{Z_3}\right)^{-1} , \quad 
\nn
%%%%%%%
&  
e^B=  Z_e \,e\,\mu^{\frac{\epsilon} {2} }\,, \quad
Z_e = \frac{Z_5 \left(\frac{Z_1}{Z_3}\right)^{1-\frac{d}{2}}
\, \left( {\frac{Z_2}{Z_3}} \right)^{-1/2}  }
{\sqrt{Z_1} \, Z_4 } \,.
\end{align}
%%%%%%%%%%%
Observing that there exists a freedom to change the renormalization of the fields and the renormalization of momenta
without affecting the action, we have exploited it by requiring $k^B_d = k_d $, which is equivalent to measuring the scaling dimensions of all the other quantities relative to the scaling dimension of $ k_d $. $S$ now represents the renormalized action (also known as the Wilsonian effective action) because it consists of the renormalized quantities.
Basically, we have written the fundamental action of our theory in two different ways \cite{srednicki}, which allows us to subtract off the divergent parts (represented by $S_{CT}$).
% ref: https://web.physics.ucsb.edu/~mark/ms-qft-DRAFT.pdf pg 178 -- Sredniki book

%%%%%%%%%%%%%%%%%%%%%%%%%%%%%%
\subsection{RG flow equations from one-loop results}
%%%%%%%%%%%%%%%%%

At one-loop order, the divergent contributions are obtained from Eqs.\eqref{eqpi} and \eqref{eqferm1}. These lead to
\begin{align}
\label{eqZvals}
Z_1 & = 1-   \frac{ e^{4/3} \, \, {\mathcal U}_1 } 
{ 2^{2/3}
\,\epsilon}  \,, \quad
%%%%%%%%%%%
Z_2 = 1 
\,,\quad
%%%%%%%%%%%
Z_3 = 1  \,,\quad 
Z_4 =1 \,,\quad Z_5 = 1 \,, \quad
%%%%%%%%%%%%%%%%
b  = \frac{ \pi^{3/4}}
{ 32 \,\sqrt{2} \,\, \Gamma^2 \big( 3/4 \big)
\,\Gamma \left(7/4\right) } \,, \quad
%%%%%%%%%%%
{\mathcal U}_1
 = \frac{\sqrt{2} \,\,
 \Gamma \big (\frac{5}{4}\big)}
 {3\, \sqrt 3 \, \pi^{7/4} \, b^{1/3}  } \,. 
\end{align}
%%%%%%%%%%%%%%%%%%%%%
To this leading order correction, we find that $Z_2 = Z_3 $, and they do not get any correction from the loop integrals.

Because $Z_{2} = Z_{3} $, we define a single dynamical critical exponent for the fermions as
\begin{align}
& z = 1 +
\frac{\partial \ln \big ( \frac{Z_1} {Z_2} \big ) }  
{\partial \ln \mu}
 = 1 +
\frac{\partial \ln \big ( \frac{Z_1} {Z_3} \big ) } 
{\partial \ln \mu}\,.
\end{align}
This applies to our one-loop level calculations where the $\delta_k $-part, as a whole, is not renormalized. Furthermore, the anomalous dimensions for the fermions and the bosons are given by
\begin{align}
\eta_\psi  = \frac{1} {2} 
\frac{\partial \ln Z_\psi }  
{\partial \ln \mu}  \text{ and }
\eta_\phi  = \frac{1} {2} \frac{\partial \ln Z_\phi }  
{\partial \ln \mu} \,,
\end{align}
respectively.
We also define the beta functions for $e$ as
\begin{align}
\beta_e  =  \frac{ d  e }  { d\ln \mu}  \,.
\end{align}

The sole purpose of the introduction of the \textit{ad hoc} mass scale $\mu $ is to regularize the theory, thus eliminating the infinities emerging from the loop integrals of Feynman diagrams. However, since physical quantities must be independent of
$\mu $, as $\mu$ is not really a parameter of the fundamental theory, the bare parameters must be independent of it as well. Imposing this condition, as well as the requirement that the regular (i.e., non-singular) parts of the final solutions are of the forms
\begin{align}
\label{eqexp}
z = z^{(0)} \,, \quad
% \tilde z = \tilde z^{(0)} \,, \quad
\eta_\psi = \eta_\psi^{ (0)}  + \eta_\psi^{ (1)} \,\epsilon\,, \quad
\eta_\phi = \eta_\phi^{ (0)}  + \eta_\phi^{ (1)} \,\epsilon\,, \quad
\beta_e = \beta _e^{ (0)}  + \beta _e^{ (1)} \,\epsilon\,,
\end{align}
in the limit $\epsilon  \rightarrow 0 $, we get the following differential equations:
%%%%%%%%%%%%%%%%%%%%%%%%%%
\begin{align}
z & = 1 + \beta_e^{(1)} \, \frac{  \partial Z_1^{(1)} }{ \partial e} \, , \quad
%%%%%%%%%%%%%%%%%%%%%%%%%%%%%%
\eta_\psi  = \frac{1} {4} 
\left(
5 -5 \,z 
+ 2\, \frac{ \partial Z_1^{(1)}  }
{\partial e} \, \beta _e^{(1)}   \right)
+ \frac{(z-1) \,\epsilon} {2} \,, \quad
%%%%%%%%%%%
\eta_\phi = \frac{ 3 -3\, z } {4} 
+ \frac{(z-1) \,\epsilon} {2}\,, \nn
%%%%%%%%%%%%%%%%%%%%%%%%%%
\frac{ 4\,\beta_e^{(0)} } {e} & =- e \,z\,
\frac{  \partial Z_1^{ (1)}} {\partial e } + z-1 \,, \quad
%%%%%%
\beta_e^{(1)} = -\,\frac{e\, z} {2}  \,.
\end{align}
The above set of equations have been obtained by (1) demanding that $\frac{ d }{d \ln \mu} \,(\mbox{bare quantity}) =0$;
(2) plugging in the values from Eqs.~\eqref{eqZvals} and \eqref{eqexp}; (3) expanding in powers of $\epsilon$;
and (4) matching the coefficients of the regular powers of $\epsilon $ in the resulting equations.
Solving the above equations, we get
%%%%%%%%%%%%%%%%%%%%%%%%%%%%
\begin{align}
 & - \frac{ \beta_e} {e}
 =\ \frac{6 \,\epsilon -3 \times 2^{1/3} \,U_1 \tilde{e}}
 {12- 2^{7/3} \, {\mathcal U}_1 \,\tilde{e}} \,,\quad
%%%%%%%%%%%%%%%%%%%%%%%%%%%%%%%%%%%%%%%%%%%%%%%%%%%%%
z  = \frac{3}{3-2^{1/3} \, {\mathcal U}_1 \,\tilde{e}} \,,\quad
%%%%%%%%%%%%%%%%%%%%%%
\eta_\psi =\eta_\phi =\frac{ (3-2 \,\epsilon )  \, {\mathcal U}_1 \,\tilde{e}}
{4 \, {\mathcal U}_1 \tilde{e}-6\times 2^{2/3}} \,,
\end{align}
%%%%%%%%%%%%%%%%%%%%%%%%%%%%%%%%%%
where
\begin{align}
\tilde e = e^{4/3} \,.
\end{align}
Since we are interested in the behaviour at the IR energy scales, we determine the RG flows with respect to the logarithmic length scale $l$, which are given by the derivative
\begin{align}
\frac{ d e}{ d l} \equiv - \,\beta_e  \,,
\end{align}
for the coupling constant $e$.

%%%%%%%%%%%%%%%%%%%%%%%%%%%%%%
\subsection{Stable fixed points}

The fixed points of the theory are obtained as the points where the beta function ($\beta_e$) goes to zero. The nomenclature originates from the fact that these are the \textit{equilibrium points} of the differential equations describing the RG flows in the space of the coupling constant $ e $. The explicit expressions for these fixed points are easily obtained as $e=0$ and $\tilde e = 2^{2/3} \, \epsilon /  {\mathcal U}_1 $.
In order to determine the stability of a fixed point, one needs to figure out whether the flow lines (in the IR), given by $\lbrace -\beta_{ e } \rbrace $, are towards or away from it. Accordingly, they are classified as stable or unstable. In the range $\epsilon \in (0, 1/2]$, we find precisely one stable (interacting) fixed point for each value of $\epsilon $, which is $ 2^{2/3} \, \epsilon /  {\mathcal U}_1$. The value of $e=0$ corresponds to an unstable Gaussian (non-interacting) fixed point. At the stable fixed point, we have
\begin{align}
z = 1+\frac{2\,\epsilon}{3} \,, \quad \eta_\phi = \eta_\psi = -\,\frac{\epsilon}{2}\,.
\end{align}
Since $Z_2 = Z_3 = 1$ at the one-loop order, we do not find any flattening of the Fermi surface at the hot-spots, unlike the interpretation of Ref.~\cite{matthias_2kf}, where the authors performed the RG using the pole of $\Sigma_2 $ at $d=3/2$.

%%%%%%%%%%%%%%%%%%%%%%%%%%%%%%%%%
\section{Discussions and outlook}
\label{secsum}

In this paper, we have revisited the QFT of the quantum critical point emerging at the continuous phase transition from a normal metal phase to an ordered phase involving an incommensurate CDW modulation. In our one-loop computations, we have used a dressed boson propagator by including the Landau-damping correction $\Pi_{\rm LD}$, which is instrumental in inducing the non-Fermi liquid behaviour with a characteristic scaling of $\text{sgn} (k_0) |k_0|^{2/3} $ for the fermion self-energy \cite{denis, ips-uv-ir1, ips-uv-ir2, ips-fflo, ips-nfl-u1, ips-rafael, peng_cavity}.
%%%%%%%%%%%%%%%%%%%%%%%%%%%%%%%%%
In Ref.~\cite{matthias_2kf}, the authors computed the fermion self-energy $\Sigma (k) $ by assuming the \textit{ad hoc} form of the boson-self energy $\tilde \Pi(k)$ to be $\propto  - k_{d - 1}$, which they argued is generated in RG due to symmetry arguments. However, a careful analysis shows that the RG procedure
%%%%%%%%%%%%%%%%
does not generate any relevant divergent term proportional to $e_k $ (or, $ k_{d - 1}$ for that matter) in the IR limit [cf. Eq.~\eqref{eqpi}]. This conclusion is also physically sensible because a term proportional to $e_k $ or $ k_{d - 1}$ would have implied that the bosonic energy is unbounded from below (which is unphysical).
 %%%%%%%%%%%%%%%%%%
As a consequence of not including the all-important Landau-damped term in the dressed bosonic propagator, they did not obtain any contribution in $\Sigma (k) $ proportional to $\mathbf \Gamma \cdot \mathbf K$ and, hence, missed the crucial $\text{sgn} (k_0) |k_0|^{2/3} $-dependence at $d = 2$. 
They concluded that the correct frequency-dependence would show up at two-loop order. 
Furthermore, Halbinger \textit{et al.} included the part proportional to $ \gamma_{d-1} $ in the one-loop fermion self-energy in the counterterms, which leads to their conclusion of the flattening of the Fermi surface at the hot-spots (as found in the RPA calculations of Ref.~\cite{metzner2}). However, that term has a factor of $\Gamma(3/2 - d)$, which first blows up at $d=3/2$ --- hence, it should not contribute to the beta-functions for the flows towards IR, when $d_c$ is equal to $5/2$. 
Their integrals are actually somewhat similar to the scenario for our $ I_3 $ calculation, and we have argued in detail why that term should not be included while computing the RG flows. 

While we have addressed here the behaviour of a system with two hot-spots in the incommensurate CDW setting in two spatial dimensions, it will be worthwhile to extend it to the case of three dimensions \cite{cdw_3d}, where we expect a marginal Fermi liquid behaviour, analogous to the results found in Refs.~\cite{ips-uv-ir1, ips-uv-ir2, ips-nfl-u1}. Another interesting direction to investigate is the scenario when the Fermi surface harbours two pairs of hot-spots \cite{metzner3}. Last, but not the least, we would like to compute the nature of superconducting instabilities in the presence of these critical CDW bosons, utilizing the RG set-ups constructed in Refs.~\cite{ips-sc, ips-qbt-sc}.

%%%%%%%%%%%%%%%%%%%%%%%%%%%%%%%%%%%%%%
\section*{Acknowledgments}
We are grateful for useful discussions with Sung-Sik Lee, Ashoke Sen, and Dimitri Pimenov. We thank Peng Rao for participating in the initial stages of the calculations. This research, leading to the results reported, has received funding from the European Union's Horizon 2020 research and innovation programme under the Marie Skłodowska-Curie grant agreement number 754340.

%%%%%%%%%%%%%%
\bibliography{biblio_cdw}

\end{document}